# Structural and electronic properties of new "122" pnictogen-free superconductor $SrPd_2Ge_2$ as compared with $SrNi_2Ge_2$ and $SrNi_2As_2$: first principles calculations

**I. R. Shein\* and A. L. Ivanovskii**

*Institute of Solid State Chemistry, Ural Division, Russian Academy of Sciences, Pervomaiskaya St., 91, Ekaterinburg, 620990 Russia*

**Abstract**

**Very recently the new low-temperature ($T_C \sim 3K$) superconductor (SC) $SrPd_2Ge_2$ has been reported. This compound is isostructural with curently intensively studied group of so-called "122" SC's (based on tetragonal $AM_2Pn_2$ phases, where $A$ are Sr, Ba; $M$ are $d$ metals and $Pn$ are pnictogens: As or P), but it is pnictogen-free. Here, by means of first-principle FLAPW-GGA calculations, we have studied the electronic structure of new SC $SrPd_2Ge_2$. The band structure, total and partial densities of states and Fermi surface topology for $SrPd_2Ge_2$ are evaluated and discussed in comparison with those of isostructural $SrNi_2Ge_2$ and $SrNi_2As_2$ phases.**

\* Corresponding author
*E-mail address:* shein@ihim.uran.ru (I.R. Shein).

## 1. Introduction

Since Hosono *et al.* have found superconductivity with $T_C \sim$ 27K in iron pnictides [1], so far, six related groups of novel superconducting materials have been uncovered in this class: *ROMPn* (so-called "1111" phases), *AM$_2$Pn$_2$* (so-called "122" phases), *A*FeAs (so-called "111" phases), FeSe$_{1-x}$, and more complex $A_3M_2M'_2Pn_2O_5$ and $A_4M_2M'_2Pn_2O_6$ (so-called "32225" and "42226" (denoted sometimes also as 21113) phases, respectively). Here, *A* are alkaline metals or alkaline earth metals, *R* are rare earth metals, *M, M'* are transition metals, and *Pn* are pnictogens.

All the mentioned materials are anisotropic (quasi-two-dimensional, 2*D*) systems with a crystal structure formed by negatively charged blocks [$M_2Pn_2$]$^{\delta-}$ alternating with positively charged blocks or atomic sheets. The superconductivity in all these systems is attributed to the [$M_2Pn_2$] blocks which make a decisive contribution to the near-Fermi region of these materials, while the positively charged blocks (for example, [*RO*]$^{\delta+}$ or $A^{\delta+}$) serve as the so-called charge reservoirs, see [2-9].

In particular, a broad family of the mentioned "122" superconductors (SC's) with transition temperatures up to $T_C \sim$ 38K (see [2-5,10-16]) based on *A*Fe$_2$As$_2$ phases (*A* = Ca, Sr or Ba) was prepared by hole doping, *i.e.* by partial substitution of alkaline metals for alkaline earth metals; or by partial replacement of Fe (in [Fe$_2$As$_2$] blocks) by other 3*d* transition metals as Mn, Co or Ni. Note, that all these materials have a high content of ***magnetic metals*** (Fe, Mn, Co and Ni). Interestingly, the same result may be achieved by ***partial substitution*** of some ***non-magnetic 4d, 5d metals*** (Ru, Ir *etc.*) by ***magnetic 3d metal*** (Fe), see [17-20].

Moreover, very recently a new "122" iron-free pnictides, namely, SrNi$_2$As$_2$,[18] BaNi$_2$As$_2$, BaNi$_2$P$_2$,[20] SrRu$_2$As$_2$, BaRu$_2$As$_2$ [20] SrRh$_2$As$_2$ [21], BaRh$_2$P$_2$ ($T_C \sim$ 1K), BaIr$_2$P$_2$ and BaRh$_2$As$_2$ [22] were synthesized. These materials, where ***magnetic metal*** (Fe) ***is completely replaced by the other 3d magnetic metal (***Ni***) or non-magnetic metals*** (Ru, Rh) ***arsenic is completely replaced by other pnictogen – phosphorus***, belong to the above mentioned "122" family and are low-temperature SC's (T$_C \sim$ 0.3-3.0K), see [21-24].

Thus, the recent report [25] on the synthesis of a new low-temperature ($T_C \sim$ 3K) SC SrPd$_2$Ge$_2$ isostructural with the group of "122" SC's, ***but in contrast to these systems, is pnictogen-free***, seems very interesting.

In this paper by means of first-principle FLAPW-GGA calculations, we have investigated the structure and electronic properties of newly discovered SC SrPd$_2$Ge$_2$ in comparison with SrNi$_2$Ge$_2$ [26,27] and SrNi$_2$As$_2$. That allows us to compare the above properties of these isostructural phases as a function of the *p*-element type (Ge *versus* As, *i.e.* SrPd$_2$Ge$_2$, SrNi$_2$Ge$_2$ ↔ SrNi$_2$As$_2$) and the *d*-metal type (Pd *versus* Ni, *i.e.* SrPd$_2$Ge$_2$ ↔ SrNi$_2$Ge$_2$, SrNi$_2$As$_2$). As a result, the optimized lattice parameters, band structures, densities of states (DOSs), Fermi surfaces are presented and analyzed. The Sommerfeld constants ($\gamma$) and the Pauli paramagnetic susceptibility ($\chi$) are estimated.

## 2. Models and method

All the considered ternary "122" phases SrPd$_2$Ge$_2$, SrNi$_2$Ge$_2$ and SrNi$_2$As$_2$ crystallize in the quasi-two-dimensional ThCr$_2$Si$_2$-type tetragonal structure, space group I4/*mmm*; Z=2. The structure is built up of [$M_2X_2$] blocks (where $M$ = Pd, Ni; **X** = Ge or As) alternating with Sr atomic sheets stacked along the *z* axis, as shown on Fig. 1. The atomic positions are: Sr: 2*a* (0,0,0), *M*: 4*d* (½,0,½) and *X* atoms: 4*e* (0,0,$z_X$), where $z_X$ are internal coordinates governing the *M-X* distances and the distortion of the $MX_4$ tetrahedra around the *M* in the [$M_2X_2$] blocks.

The calculations were carried out by means of the full-potential method with mixed basis APW+lo (LAPW) implemented in the WIEN2k suite of programs [28]. The generalized gradient correction (GGA) to exchange-correlation potential in the PBE form [29] was used. The plane-wave expansion was taken to $R_{MT} \times K_{MAX}$ equal to 7, and the *k* sampling with 10×10×10 *k*-points in the Brillouin zone was used. The calculations were performed with full-lattice optimization including internal $z_X$ coordinates. The self-consistent calculations were considered to be converged when the difference in the total energy of the crystal did not exceed 0.1 mRy and the difference in the total electronic charge did not exceed 0.001 *e* as calculated at consecutive steps.

## 3. Results and discussion

### *3.1. Structural properties*.

At the first step, the optimized atomic positions and the equilibrium structural parameters for the Sr$M_2X_2$ phases are determined; the calculated values are presented in Tables 1,2 and are in reasonable agreement with the available experimental data [25-27,30].

Our results show that replacements in blocks [Pd(Ni)$_2$Ge(As)$_2$] of the *d* metal atoms (Pd ↔ Ni) or *p* elements (Ge ↔ As) lead to *anisotropic deformations* of the crystal structure caused by strong anisotropy of inter-atomic bonds in "122' phases (see also [20,24]), but the type of such deformations will be quite different. Thus, going from SrPd$_2$Ge$_2$ to SrNi$_2$Ge$_2$, *i.e.* replacing a larger Pd atom ($R^{at}$ = 1.37 Å) by a smaller Ni atom ($R^{at}$ = 1.24 Å), the parameter *a* decreases (of about ~ 0.26 Å) whereas the inter-layer distance (parameter *c*) slightly grows. On the contrary, when going from SrNi$_2$Ge$_2$ to SrNi$_2$As$_2$, *i.e.* replacing larger As atom ($R^{at}$ = 1.48 Å) by a smaller Ge atom ($R^{at}$ = 1.39 Å), the parameter *a* slightly grows, whereas the parameter *c* decreases (at about 0.14 Å), see Table 1.

Thus, the results obtained reveal that various atomic replacements inside "122" phases can lead to essential anisotropic changes of their structural properties and are favorable for fine tuning of their geometry - in particular, in view of sensitivity of superconductivity to structural changes, see [31,32].

### *3.2. Electronic properties*.

Figures 2 and 3 show the band structures, Fermi surfaces, total and atomic-resolved *l*-projected DOSs for the Sr$M_2X_2$ phases as calculated for equilibrium geometries. We find, that total energy calculations for magnetic (in assumption of ferromagnetic spin ordering) and nonmagnetic (NM) states for these phases show that the NM state for all Sr$M_2X_2$ phases is energetically most favorable.

We discuss the most important features of electronic structure for Sr$M_2X_2$ phases, focusing on the near-Fermi region, Fig. 2. For $A$Fe$_2$As$_2$ phases, the 2$D$-like Fe $d_{xy}$, $d_{x^2-y^2}$ bands, crossing the Fermi level (E$_F$) show low $k_z$ dispersion. These bands form typical for FeAs SC's Fermi surfaces (FS's) which composed from cylinder-like electron and hole pockets, directed along $k_z$, see reviews [2-4].

The examined Sr$M_2X_2$ phases have the increased number of valence electrons (nve) as compared with FeAs-based materials ($A$Fe$_2$As$_2$ (nve = 28 e) → SrPd$_2$Ge$_2$ (nve = 30 e) → SrNi$_2$Ge$_2$ (nve = 30 e) → SrNi$_2$As$_2$ (nve = 32 e). This leads to shifting up of the Fermi level into the upper manifold of $Md$ – like bands with higher $k_z$, where in addition to $M$ $d_{xy}$ and $d_{x^2-y^2}$ orbitals, there are large contributions from $M$ $d_{xz}$, $d_{yz}$, and $d_z^2$ orbitals and $X$ $p$ states. In result, the 2$D$-like FS topology for $A$Fe$_2$As$_2$ SCs in a case of examined Sr$M_2X_2$ phases is transformed into multi-sheet three-dimensional structure, see Fig. 2. In turn, the main differences of FSs for Ge-containing "122" phases from that of SrNi$_2$As$_2$, are in the form of hole sheets which for SrPd$_2$Ge$_2$ and SrNi$_2$Ge$_2$ adopt complex 3$D$ topology. For SrNi$_2$As$_2$ the Fermi surface contains the closed hole pockets, centered at the point Z.

Comparison of the DOSs profiles for isostructural SrPd$_2$Ge$_2$, SrNi$_2$Ge$_2$ and SrNi$_2$As$_2$ allows us to understand the main differences of their valence spectra, Fig. 3. For SrPd$_2$Ge$_2$ and SrNi$_2$Ge$_2$ the occupied Ge 4$s$, 4$p$ states are placed in the energy intervals from -10.4 to -7.6 eV and from -4.9 eV to E$_F$, whereas for SrNi$_2$As$_2$ the occupied As 4$s$, 4$p$ states - in the intervals from -13.0 eV to -10.6 eV and from -5.6 eV to E$_F$, respectively.

For all Sr$M_2X_2$ phases, the valence bands (located in the intervals from -6.0 eV to E$_F$) are formed predominantly by mixed Pd(Ni) $d$ – Ge(As) $p$ states which are responsible for the covalent Pg(Ni)-Ge(As) bonding inside [Pd(Ni)$_2$Ge(As)$_2$] blocks. On the contrary, the contributions from the valence states of Sr to these bands are negligible, *i.e.* in Sr$M_2X_2$ phases these atoms are in the form of cations Sr$^{2+}$ and act as the electronic donors. Thus, SrPd$_2$Ge$_2$, SrNi$_2$Ge$_2$ and SrNi$_2$As$_2$ phases may be described by two types of oppositely charged blocks: …[(Pd,Ni)$_2$(Ge,As)$_2$]$^{2-}$/Sr$^{2+}$/[(Pd,Ni)$_2$(Ge,As)$_2$]$^{2-}$/Sr$^{2+}$…. The [(Pd,Ni)$_2$(Ge,As)$_2$] block is characterized by mixed covalent-ionic-metallic Pd(Ni)- Ge(As) bonding due to the hybridization of Pd(Ni) 4$d$ – Ge(As)$p$ states,
Pd(Ni) M → Ge(As) charge transfer and delocalization of near-Fermi Pd(Ni) $d$ states, respectively. Between the adjacent [(Pd,Ni)$_2$(Ge,As)$_2$] blocks and $A$ atomic sheets, there is ionic bonding due to $A^{2+}$ → [(Pd,Ni)$_2$(Ge,As)$_2$]$^{2-}$ charge transfer.

Since the near the Fermi surface electrons are involved in the formation of superconducting state, it is important to figure out their nature. The total and orbital decomposed partial DOSs at the Fermi level, N(E$_F$), are shown in Table 3. For all Sr$M_2X_2$ phases, the main contributions to N(E$_F$) come from the $M$ $d$ states, whereas the contributions from Ge(As) states are much smaller. According to our estimations, the values of the N(E$_F$) as well as the Sommerfeld constants ($\gamma$) and the Pauli paramagnetic susceptibility ($\chi$) (obtained under the assumption of the free electron model as $\gamma$ =

$(\pi^2/3)N(E_F)k_B^2$ and $\chi = \mu_B^2 N(E_F)$, Table 3) decrease in the following sequence: SrNi$_2$Ge$_2$ > SrNi$_2$As$_2$ > SrPd$_2$Ge$_2$. Due to the reduction of N(E$_F$), the observed lowering of transition temperature for SrNi$_2$As$_2$ ($T_C \sim 0.6$K [30]) as compared to recently discovered SC SrPd$_2$Ge$_2$ ($T_C \sim 3.0$K [25]) cannot be explained in the framework of conventional electron-phonon BCS theory in terms of the electronic factor. As is known, in the BCS strong coupling limit, $T_C$ is expressed by the McMillan formula: $T_C \sim <\omega> \exp\{f(\lambda)\}$, where $<\omega>$ represents an averaged phonon frequency, the coupling constant $\lambda = N(E_F) <I^2>/<M\omega^2>$, where $<I^2>$ is an averaged electron-ion matrix element squared and M is an atomic mass. Therefore, the reasons why the $T_C$ for SrPd$_2$Ge$_2$ is higher than for SrNi$_2$As$_2$ can be connected to the phonon system. Indeed, as going from SrNi$_2$As$_2$ to SrPd$_2$Ge$_2$, the lattice parameter *a*, which defines the intra-atomic distances inside the conducting blocks [Pd(Ni)$_2$Ge(As)$_2$], grows noticeably. This can lead to softening of phonon modes and promote growth of the coupling constant $\lambda$. Certainly, to check this assumption one need to calculate directly the phonon spectra for these phases.

Finally, we note, that superconductivity for SrNi$_2$Ge$_2$ has been not yet reported, contrary to SrPd$_2$Ge$_2$ (T$_C \sim 3$K [25]) and SrNi$_2$As$_2$ (T$_C \sim 0.6$K [30]). The data obtained here allow us to assume the emergence of low-temperature superconductivity for SrNi$_2$Ge$_2$, and we believe that related experiments will be of high interest.

## 4. Conclusions

We studied structural and electronic properties of recently synthesized low-temperature ($T_C \sim 3$K) superconductor SrPd$_2$Ge$_2$, belonging to the intensively studied group of "122" FeAs SC's, but pnictogen-free. The obtained results were compared with properties of isostructural SrNi$_2$Ge$_2$ and SrNi$_2$As$_2$ phases.

Our results show that the atomic replacements in these phases: (Ga ↔ As) or (Pd ↔ Ni) lead to various types of *anisotropic deformations* of the crystal structures and are useful for purposeful tuning of their geometries, in particular, due to known sensitivity of superconductivity to structural changes for "122" materials.

Our band structure calculations for SrPd$_2$Ge$_2$, SrNi$_2$Ge$_2$ and SrNi$_2$As$_2$ show that these materials are metals with dispersive bands at the Fermi level. The near-Fermi valence bands in these phases are derived mainly from (Pd,Ni) *d* states. As distinct from *A*Fe$_2$As$_2$ phases, the Fermi level in SrPd$_2$Ge$_2$, SrNi$_2$Ge$_2$ and SrNi$_2$As$_2$ phases is shifted from a manifold of quasi-two-dimensional low-dispersive *d* - like bands to the upper bands with higher dispersion E(*k*) as a result of increased electron concentration. In result the Fermi surfaces for these materials differ essentially from the FSs of *A*Fe$_2$As$_2$ phases and adopt a multi-sheet three-dimensional type.

According to our calculations, the observed lowering of transition temperature for SrNi$_2$As$_2$ as compared with recently discovered SC SrPd$_2$Ge$_2$ cannot be explained in terms of the electronic factor but migth be connected to the phonon system.

The SrNi$_2$Ge$_2$ phase may be expected as a possible superconductor, and it will be of interest to probe experimentally this material.

**Acknowledgments**
Financial support from the RFBR (Grant 09-03-00946-a) is gratefully acknowledged.

**Table 1.**
The optimized atomic positions and internal coordinates ($z_X$) for $SrPd_2Ge_2$, $SrNi_2Ge_2$ and $SrNi_2As_2$ in comparison with available experiments.

| system (atomic position) | $SrPd_2Ge_2$ | $SrNi_2Ge_2$ | $SrNi_2As_2$ |
|---|---|---|---|
| Sr (2*a*) | (0,0,0) | (0,0,0) | (0,0,0) |
| Pd(Ni) (4*d*) | (0, ½, ¼) | (0, ½, ¼) | (0, ½, ¼) |
| Ge(As) (4*e*) | (0, 0, *z* ); *z* = 0.37030 (*z* = 0.37034 [25]) | (0, 0, *z* ); *z* = 0.36007 ( *z* = 0.362 [27]) | (0, 0, *z* ); *z* = 0.3621 (*z* = 0.3634 [30]) |

* available experimental data are given in parentheses.

**Table 2.**
The optimized lattice parameters (*a* and *c*, in Å) for $SrPd_2Ge_2$, $SrNi_2Ge_2$ and $SrNi_2As_2$ in comparison with available experiments.

| phase/parameter | $SrPd_2Ge_2$ | $SrNi_2Ge_2$ | $SrNi_2As_2$ |
|---|---|---|---|
| *a* | 4.4588 (4.4088 [25]) | 4.1968 (4.188 [26], 4.17 [27]) | 4.166 (4.137 [30]) |
| *c* | 10.3213 (10.1270 [25]) | 10.3595 (10.254 [26], 10.25 [27]) | 10.502 (10.188 [30]) |
| *c/a* | 2.3148 (2.297 [8]) | 2.4684 (2.448 [26], 2.458 [27]) | 2.521 (2.463 [30]) |

* available experimental data are given in parentheses.

**Table 3.**
Total ($N^{tot}(E_F)$) and partial ($N^l(E_F)$) densities of states at the Fermi level (in states/eV·atom), electronic heat capacity $\gamma$ (in mJ·K$^{-2}$·mol$^{-1}$) and molar Pauli paramagnetic susceptibility $\chi$ (in $10^{-4}$ emu/mol) for $SrPd_2Ge_2$, $SrNi_2Ge_2$ and $SrNi_2As_2$ in comparison with available experiments.

| phase/parameter | Sr *s+p* | Ge(As)*s* | Ge(As)*p* | Ge(As)*d* | Pd(Ni) 4*s* |
|---|---|---|---|---|---|
| $SrPd_2Ge_2$ | 0.023 | 0.051 | 0.359 | 0.018 | 0.065 |
| $SrNi_2Ge_2$ | 0.029 | 0.041 | 0.243 | 0.016 | 0.064 |
| $SrNi_2As_2$ | 0.013 | 0.037 | 0.274 | 0.012 | 0.029 |
| phase/parameter | Pd(Ni) 4*p* | Pd(Ni) 3*d* | total | $\gamma$ | $\chi$ |
| $SrPd_2Ge_2$ | 0.126 | 0.914 | 2.902 | 6.84 | 0.93 |
| $SrNi_2Ge_2$ | 0.137 | 1.603 | 3.332 | 7.85 | 1.07 |
| $SrNi_2As_2$ | 0.095 | 1.676 | 3.173 | 7.48 (8.7 [30]) | 1.02 |

* available experimental data are given in parentheses.

**Figures**

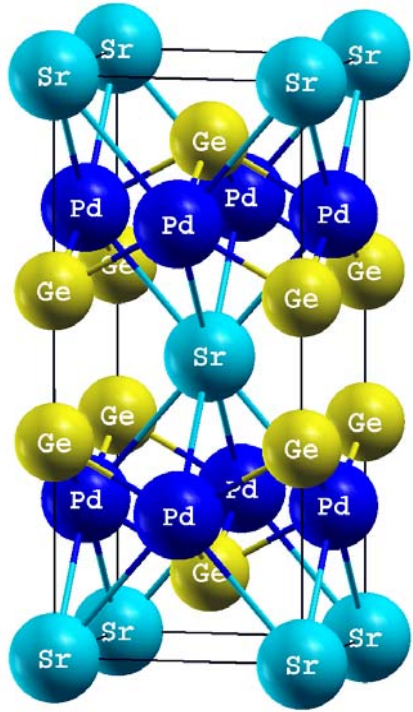

Fig. 1. Crystal structure of tetragonal (ThCr$_2$Si$_2$-like) SrPd$_2$Ge$_2$ phase. [Pd$_2$Ge$_2$] blocks and Sr atomic sheets are stacked along the *c* axis.

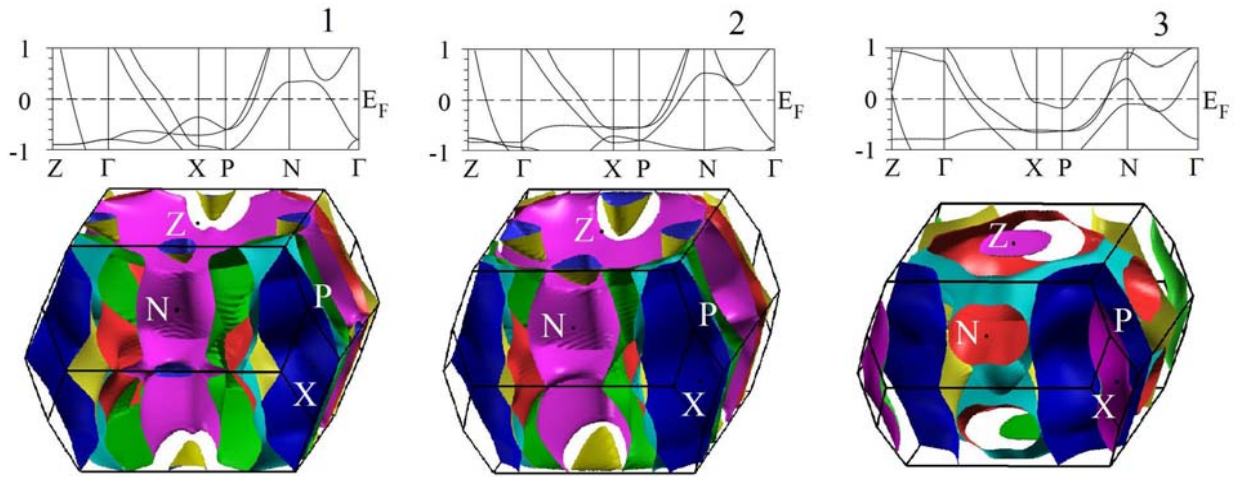

Fig. 2. Near-Fermi electronic bands and Fermi surfaces for SrPd$_2$Ge$_2$ (1), SrNi$_2$Ge$_2$ (2) and SrNi$_2$As$_2$ (3).

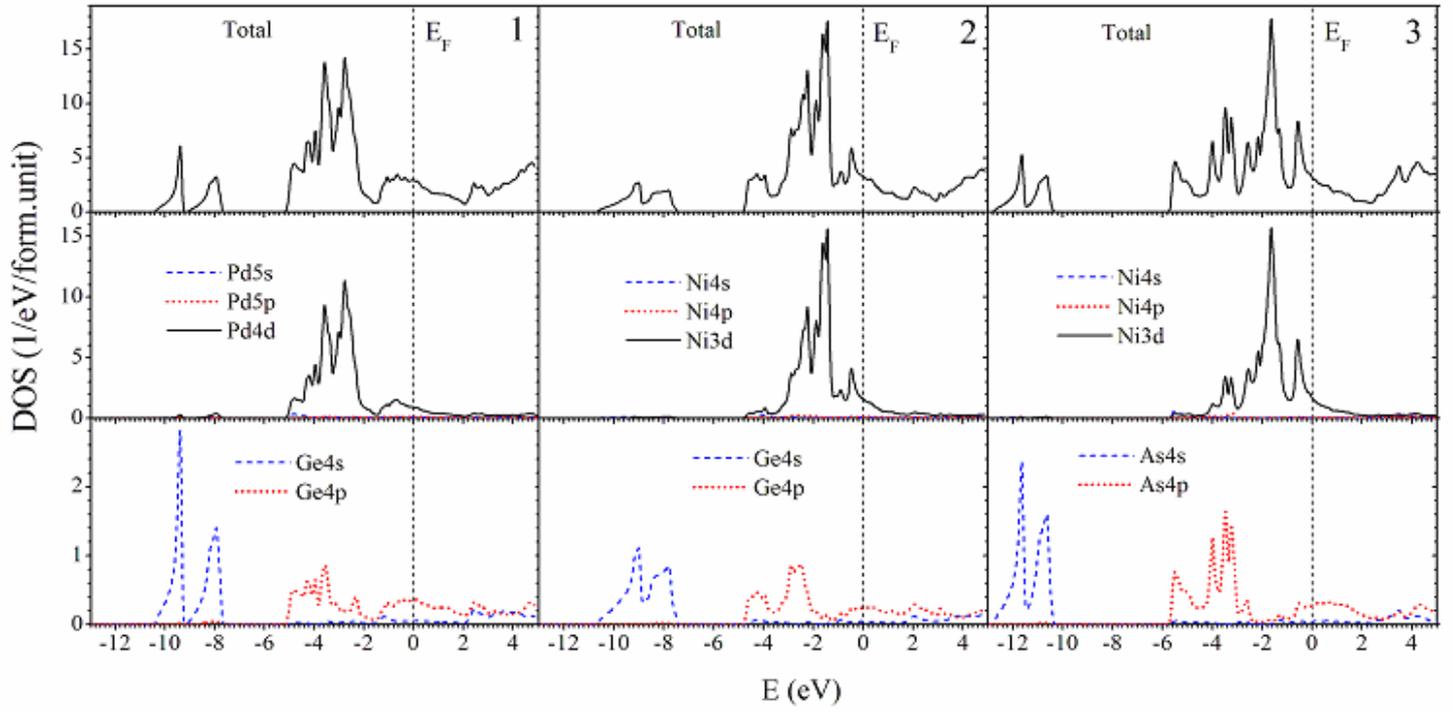

Fig. 3. Total and partial densities of states for SrPd$_2$Ge$_2$ (1), SrNi$_2$Ge$_2$ (2) and SrNi$_2$As$_2$ (3).